\documentclass[aps,prb,twocolumn,amssymb,amsmath,showkeys,showpacs,superscriptaddress]{revtex4-1}
\usepackage{bm}
\usepackage{times}
\usepackage{graphicx}
\usepackage{color}
\usepackage{dcolumn}
\usepackage[colorlinks=true, pdfstartview=FitV, linkcolor=blue, citecolor=blue, urlcolor=blue]{hyperref}
\usepackage[normalem]{ulem}
\usepackage{lipsum}
\usepackage{makecell}

\begin{document}
\title{Bulk-boundary-transport correspondence of the second-order topological insulators}
\author{Yuxiong Long$^\S$}
\affiliation{College of Physics and Optoelectronic Engineering, Shenzhen University, Shenzhen 518060, China}
\author{Miaomiao Wei$^\S$}
\affiliation{College of Physics and Optoelectronic Engineering, Shenzhen University, Shenzhen 518060, China}
\author{Fuming Xu}
\email[]{xufuming@szu.edu.cn}
\affiliation{College of Physics and Optoelectronic Engineering, Shenzhen University, Shenzhen 518060, China}
\author{Jian Wang}
\email[]{jianwang@hku.hk}
\affiliation{College of Physics and Optoelectronic Engineering, Shenzhen University, Shenzhen 518060, China}
\affiliation{Department of Physics, The University of Hong Kong, Pokfulam Road, Hong Kong, China}
\affiliation{Department of Physics, University of Science and Technology of China, Hefei, Anhui 230026, China}

\begin{abstract}
The bulk-boundary correspondence of the second-order topological insulator (SOTI) has been well established, but a universal transport signature for open systems is still absent. For a variety of SOTIs induced by applying in-plane magnetic fields in Z$_2$-invariant first-order TIs, rotating this magnetic field features the spin pump mechanism while maintaining the SOTI phase. We demonstrate that, this spin pump can generate quantized pure spin current when tuning the magnetic field strength, which corresponds to the formation of topological corner states characterizing SOTI in two-dimensional (2D) systems. Quantized spin pump is discovered in various 2D and 3D SOTI models evolved from Z$_2$-invariant TIs, which is robust against disorder and universally independent of system parameters including Fermi energy, system size, magnetic field strength, and pumping frequency. These findings suggest that this universal quantized spin pump can characterize the bulk-boundary-transport correspondence of SOTIs. Quantized spin pump can also be realized by combining pseudo spin such as the orbital degree of freedom with the rotating magnetic field, which could be achieved in higher-order photonic or acoustic topological systems. Such a quantized spin pump is promising as an accurate and stable single-spin source.
\end{abstract}

\maketitle

\section{Introduction}
Topological phases of matter have corresponding signatures in various spatial domains. For instance, Chern insulator is characterized by Chern number in bulk systems; when it is confined with boundaries, spinless edge states emerge; transport measurements show universal quantized Hall conductance in open systems\cite{Klitzing,Overney}. Such bulk-boundary-transport correspondence is also valid for topological insulators (TI, or quantum spin Hall (QSH) effect)\cite{Kane,BHZ,Konig,SCZhang09,QXue10}, which is represented by the Z$_2$ invariant, helical edge states, and universal quantized spin Hall conductance, respectively. Universal quantized transport signature is the manifestation of bulk topological invariants in open systems\cite{AHE2010,Niu2010,Qi2011}.

For higher-order TIs, the bulk-boundary correspondence has been well established. The second-order TI (SOTI) is described by topological invariants including quadruple moment\cite{Benalcazar,Shen20,Duan21}, Wannier center\cite{Ezawa18PRB}, nested Wilson loop\cite{Benalcazar,Fulga18,Cho19}, Zak phase\cite{Liu21}, winding number\cite{Park19,Ren}, etc. Meanwhile, the boundary states of SOTI are characterized by 0-dimensional (0D) corner states in 2D systems and 1D hinge modes in 3D systems\cite{Sheng19,Schindler,Chen20Hinge,HOTIRev1}. Topological corner states of SOTI are localized in real space and the corresponding energy levels are embedded in a large energy gap, which hardly contribute to transport. There are a few attempts on the resonant tunneling\cite{KTWang2021,KTWang2022} and charge pumping\cite{KTWang2022,HJiang2022} properties of topological corner states. However, as far as we know, universal quantized transport signature of SOTI is absent so far.

Thouless pump\cite{Thouless} can reveal the band topology and describe quantized charge pumping in bulk higher-order TIs\cite{Grusdt2022,ThoulessRev}, but not for open systems. On the other hand, quantum parametric pump\cite{Buttiker94,Brouwer1998} generates nonzero charge current at zero bias voltages, by adiabatically manipulating system parameters such as gate voltage or magnetic field. It was shown that the pumped charge and current in parametric pump are closely related to Berry phase and Berry curvature\cite{Sadun2000}, which suggests that parametric pump could feature quantized transport of SOTI. Specifically, when the pump is realized by periodically rotating the magnetic field, a spin pump is constructed, which induces pure dc spin current\cite{SpinPumpExpt,BauerPRL02,BauerPRB02,JWang2003}.

A variety of second-order TIs are induced by applying in-plane magnetic fields in the Z$_2$-invariant first-order TIs\cite{Song,Ezawa1,Ren,Chen1,Hu,Chen2,Zeng}. The role of this magnetic field is breaking the time-reversal symmetry; its direction can be arbitrary and rotating this in-plane magnetic field does not affect the SOTI phase. Since the magnetic field usually couples with spin, it is natural to construct spin pump via rotating the in-plane magnetic field to investigate spin transport of open SOTI systems. Here an open system means attaching leads to the SOTI system to measure its transport properties.

In this work, we demonstrate that, for SOTIs evolved from the Z$_2$-invariant TI by applying an in-plane magnetic field, integer spin quanta are pumped out per cycle when rotating this magnetic field, which gives rise to the quantization of spin transport. We show that, in finite/closed systems, the onset of quantization of spin current corresponds to formation of topological corner states representing SOTI, suggesting that quantized spin pump can serve as the quantized transport signature of SOTI that evolves from Z$_2$-invariant TIs. Quantized spin current is discovered in various 2D and 3D SOTI systems including the Bernevig-Hughes-Zhang (BHZ) model\cite{Song,Hu}, the Kane-Mele model\cite{Ren}, and the pseudo spin model\cite{Song}, where it is found to be robust against disorder and independent of system parameters such as Fermi energy, magnetic field strength and rotating frequency, and system size. Therefore, this quantized spin pump represents universal quantized transport signature and can be adopted to describe the bulk-boundary-transport correspondence of SOTIs.

\section{Spin pump via rotating magnetic field}
\begin{figure}[tbp]
\centering
\includegraphics[width=\columnwidth]{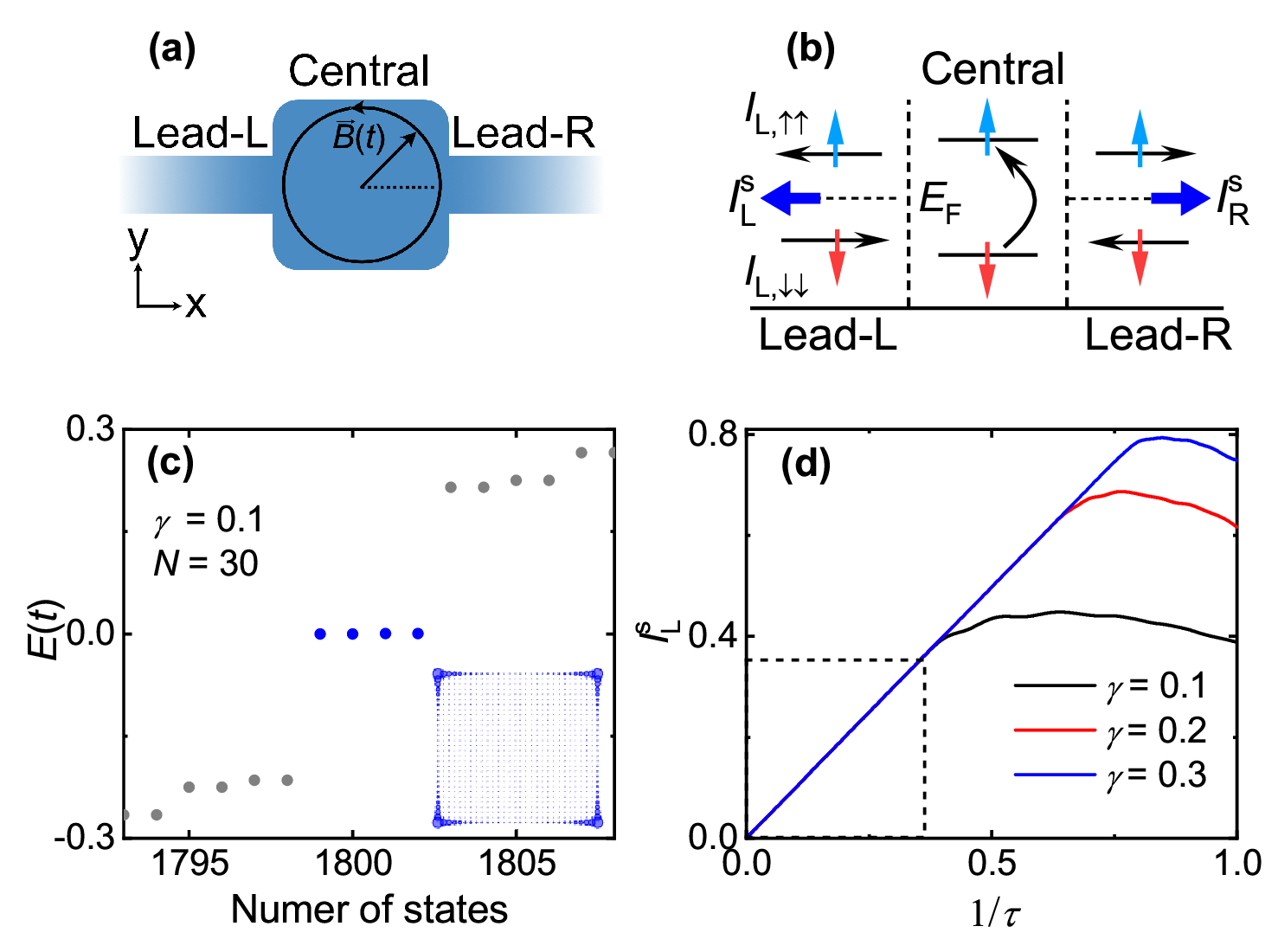}
\caption{(a) Schematic of a 2D spin pump setup. An in-plane rotating magnetic field is applied in the central region of a two-lead system with zero bias voltages. (b) Illustration of the spin pumping mechanism. (c) Energy levels for a $N \times N$ lattice of the BHZ model of SOTI. Zero energy levels (blue dots) corresponds to topological corner states which locates at the four corners of the square lattice shown in the inset. (d) Pumped spin current in the left lead as a function of the pumping frequency ($1/\tau = \omega/2\pi$) for different magnetic field strengths.}\label{fig1}
\end{figure}

We start with a 2D model system for spin pump, which is shown in Fig.~\ref{fig1}(a). A rotating magnetic field is applied in the central region of this two-lead system, whose Hamiltonian consists of three parts:
\begin{equation}
H = H_0 + H_{rot} + H_T.
\end{equation}
$H_0$ is the Hamiltonian of the central region as well as the left and right leads, which is assumed to preserve time-reversal symmetry (TRS) and spin degeneracy. $H_{rot}$ is the time-dependent Hamiltonian induced by the rotating magnetic field in the central region, which is typically defined as
\begin{equation} \label{Hrot}
H_{rot}=\sum_i \gamma_i\left[e^{-i \omega t} c_{i \uparrow}^{\dagger} c_{i \downarrow}+e^{i \omega t} c_{i \downarrow}^{\dagger} c_{i\uparrow}\right],
\end{equation}
where $c_{i \uparrow}^{\dagger}$ ($c_{i \downarrow}$) is the creation (annihilation) operator for a spin-up $\uparrow$ (spin-down $\downarrow$) electron at site $i$ of the central region. $\gamma_{i}$ is the magnetic field strength at site $i$, which is uniform in this work. $\omega$ is the rotating frequency, which determines the pumping period $\tau = 2\pi / \omega$. Clearly, $H_{rot}$ reflects the spin flipping process through the coupling between the electron spin and the external magnetic field. $H_T$ describes the coupling between the central region and two leads. The specific form of $H_T$ depends on the Hamiltonians of the model system. Ref.~[\onlinecite{KTWang2021}] shows $H_T$ for a 2D honeycomb lattice with nearest- and next-nearest-neighbor hoppings.

The spin pumping mechanism is shown in Fig.~\ref{fig1}(b), where a spin-down electron is driven out of equilibrium by the rotating magnetic field and turned into a spin-up electron with higher energy through the spin flipping process. When periodically rotating the magnetic field, spin-down polarized currents continuously flow into the central region, and spin-up polarized currents flow out to both leads. If $H_0$ is spin conserved, charge currents cancel with each other and pure spin currents are generated in the leads. Notice that both the magnitude and directions of the spin currents persist during the pumping period, which features constant dc spin currents. Since spin currents in both leads flow out or in simultaneously, this spin pump forms a unipolar spin battery\cite{Brataas02,Sun03,JWang2003}.

The pumped spin-polarized current in lead $\alpha$ ($\alpha=L,R$) with spin $\sigma$ is obtained by the nonequilibrium Green's function formalism ($\hbar=e=1$)\cite{JWang2003,Zheng03}:
\begin{equation} \label{eq1}
\begin{split}
I_{\alpha, \sigma \sigma}&=\int \frac{d E}{2 \pi}(f(E)-f(E+\bar{\sigma} \omega)) \\
& {\mathrm{Tr}}\left[G_{\sigma \bar{\sigma}}^r(E) \Gamma_{\bar{\sigma} \bar{\sigma}}(E+\bar{\sigma} \omega) G_{\bar{\sigma} \sigma}^a(E) \Gamma_{\alpha, \sigma \sigma}(E) / 4 \pi^2\right],
\end{split}
\end{equation}
where $f(E)$ is the Fermi distribution function with energy $E$. $\sigma$/$\bar{\sigma}$ labels spin polarization $\uparrow$/$\downarrow$ in the $z$-direction and stands for $\pm 1$ for the electron energy. $\Gamma_\alpha$ is the linewidth function of lead $\alpha$, and $\Gamma = \sum_\alpha \Gamma_\alpha$. $G^{r(a)}_{\sigma\bar{\sigma}}$ is the retarded (advanced) nonequilibrium Green's function of the system. Its definition and the derivation of Eq.~(\ref{eq1}) are presented in the {\color{blue}Supporting Information}. When the Hamiltonian $H_0$ is spin conserved, this spin pump generates pure spin current,
\begin{equation}
I_\alpha^s=\frac{1}{2}\left(I_{\alpha, \uparrow \uparrow}-I_{\alpha, \downarrow \downarrow}\right).
\end{equation}
If $H_0$ is not spin conserved, only spin-polarized currents are pumped out. In the following, we investigate the spin pumping characteristics of various SOTI models.

\section{The BHZ model of SOTI}
The BHZ model has been widely used for describing quantum spin Hall states of HgTe/CdTe quantum wells\cite{BHZ,Konig} as well as topological surface states of 3D TI Bi$_2$Se$_3$\cite{SCZhang09,QXue10}. The BHZ model can also be generalized to investigate SOTI, which is achieved by introducing an in-plane magnetic field\cite{Hu,Song}. When rotating this magnetic field to facilitate spin pump, the corresponding Hamiltonians read\cite{Hu,Song}
\begin{equation}
    \begin{split}
    H_0 =\left[-m+\left(t \cos k_x+t \cos k_y\right)\right] \tau_z \otimes \sigma_z + \\
         \lambda \sin k_x \tau_x \otimes \sigma_z-\lambda \sin k_y \tau_y \otimes \sigma_z, \nonumber
    \end{split}
\end{equation}
\begin{equation}
H_{rot} = \gamma\left(\cos k_x-\cos k_y\right) \tau_0 \otimes\left(\sigma_x \cos \omega t+\sigma_y \sin \omega t\right), \nonumber
\end{equation}
where $\tau_{x/y/z}$ represent orbital degree of freedom, and $\sigma_{x/y/z}$ are Pauli matrices for spin. $\gamma$ is the strength of the uniform magnetic field, which acts on in-plane spin $\sigma_{x/y}$. $\omega$ is the rotating frequency of the field and also serves as the pumping frequency. Other parameters of the model are conventionally defined in Refs.~\cite{Hu,Song}. Notice that $H_0$ preserves TRS and spin conservation of $\sigma_z$, while $H_{rot}$ breaks TRS.

For a finite system, its zero energy levels and eigenfunction distribution are shown in Fig.~\ref{fig1}(c). Four blue dots embedded in the large energy gap corresponds to the second-order topological corner states, whose eigenfunctions are localized at four corners of the square lattice. When rotating the magnetic field in a two-lead open system, pure spin currents are generated. The pumped spin current in the left lead is plotted in Fig.~\ref{fig1}(d), where $I^s_L$ increases linearly with the pumping frequency in a large range ($1/\tau = \omega/2\pi$), regardless of the field strength. This observation suggests the existence of an adiabatic regime in which $I^s_L$ can be scaled with the pumping frequency. From Eq.~(\ref{eq1}), when expanding the Fermi distribution function $f(E+\omega)$ to the first order in the pumping frequency $\omega$, we have $f(E)-f(E+\omega) = -f^{\prime}(E)~ \omega$. At zero temperature, $f^{\prime}(E)$ is a $\delta$-function, which leads to:
\begin{equation}
I_{\alpha, \sigma \sigma} =\frac{\omega}{2 \pi} \operatorname{Tr}\left[G_{\sigma \bar{\sigma}}^r(E)\Gamma_{\bar{\sigma} \bar{\sigma}}(E) G_{\bar{\sigma} \sigma}^a(E) \Gamma_{\alpha, \sigma \sigma}(E) / 4 \pi^2\right]. \nonumber
\end{equation}
This expression shows that $I_{\alpha, \sigma \sigma}$ scales linearly with $\omega$ in a certain range, where the scaled current $2 \pi I_{\alpha, \sigma \sigma}/\omega= \tau I_{\alpha, \sigma \sigma}$ is a constant.

\begin{figure}[tbp]
\centering
\includegraphics[width=\columnwidth]{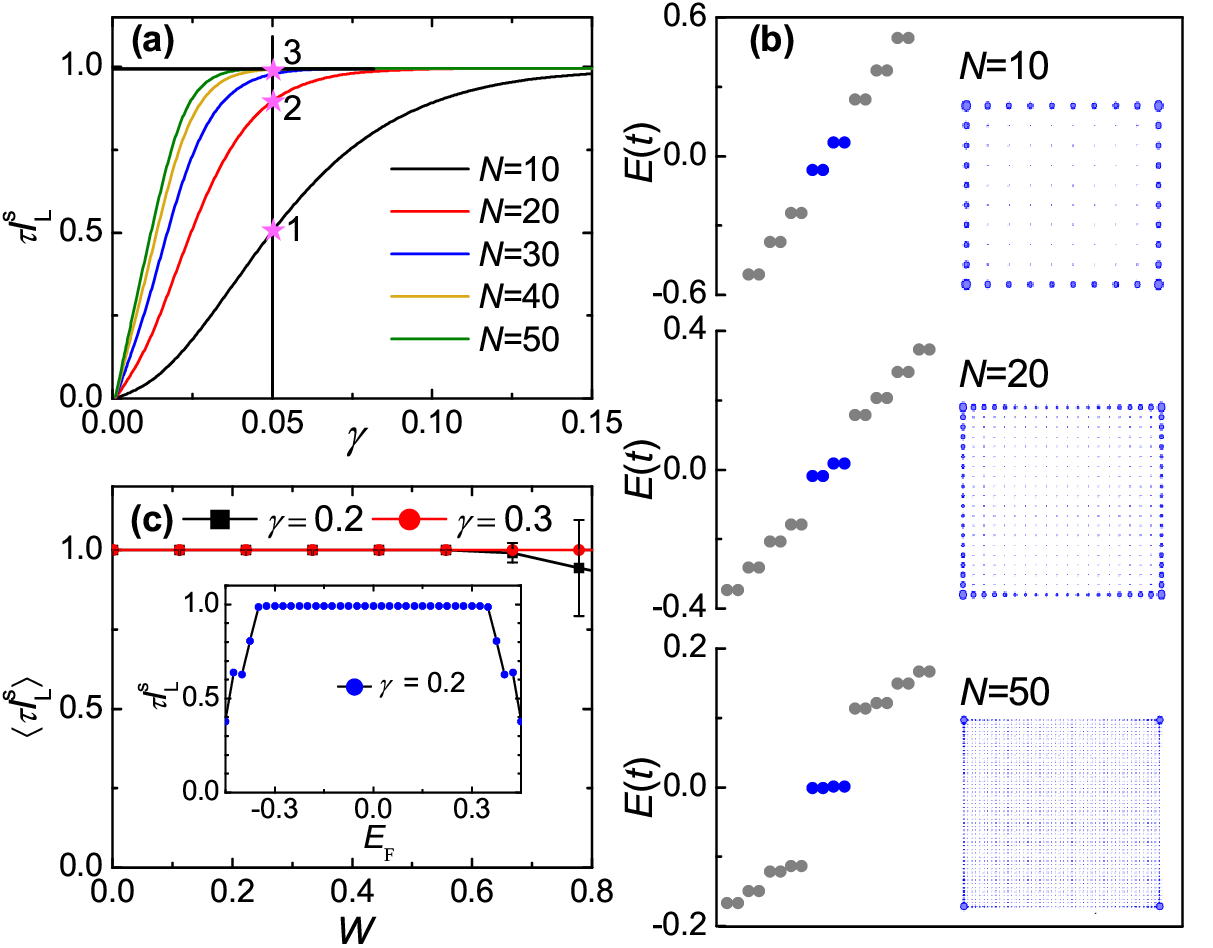}
\caption{(a) $\tau I^s_L$ versus $\gamma$ for different system sizes. (b) Energy levels near $E=0$ (blue dots) and their eigenfunction distribution at $\gamma=0.05$ for different system sizes $N=10, 20$, and 50, respectively. (c) Ensemble-averaged total spin current $\langle \tau I^s_L \rangle$ against disorder strength $W$. Error bar labels standard deviation. Inset shows $\tau I^s_L$ versus the Fermi energy. $N=20$ in panel (c).}\label{fig2}
\end{figure}

We further study the scaled current $\tau I^s_L$, which also represents the total spin current in one pumping period. $\tau I^s_L$ as a function of the magnetic field strength is shown in Fig.~\ref{fig2}(a). It is found that $\tau I^s_L$ increases with the strength $\gamma$, and reaches a quantized plateau beyond a critical threshold $\gamma_c$. The threshold $\gamma_c$ for quantized $\tau I^s_L$ is smaller for larger system sizes. At a fixed $\gamma$, we plot the energy levels and corresponding eigenfunction distribution for several finite systems in Fig.~\ref{fig2}(b), which are labeled by $1$ to $3$ in Fig.\ref{fig2}(a). For $N=10$, $\tau I^s_L \approx 0.5$ is far from quantization; energy levels of the finite system are evenly separated; the eigenfunctions for near zero energy levels (blue dots) uniformly spread at the boundaries of the square lattice characterizing quantum spin Hall edge states. When $N=20$, the scaled current increases to $\tau I^s_L \approx 0.9$; near zero energy levels (blue dots) deviate from other levels and the eigenfunction distribution is obviously more condense at four corners. For $N=50$, $\tau I^s_L$ is well quantized; zero energy levels are isolated in a large gap and the eigenfunction forms localized topological corner states. The numerical evidence suggests that the onset of quantized spin current in open systems corresponds to the formation of topological corner states in confined/finite systems.

\begin{figure}[bp]
\centering
\includegraphics[width=\columnwidth]{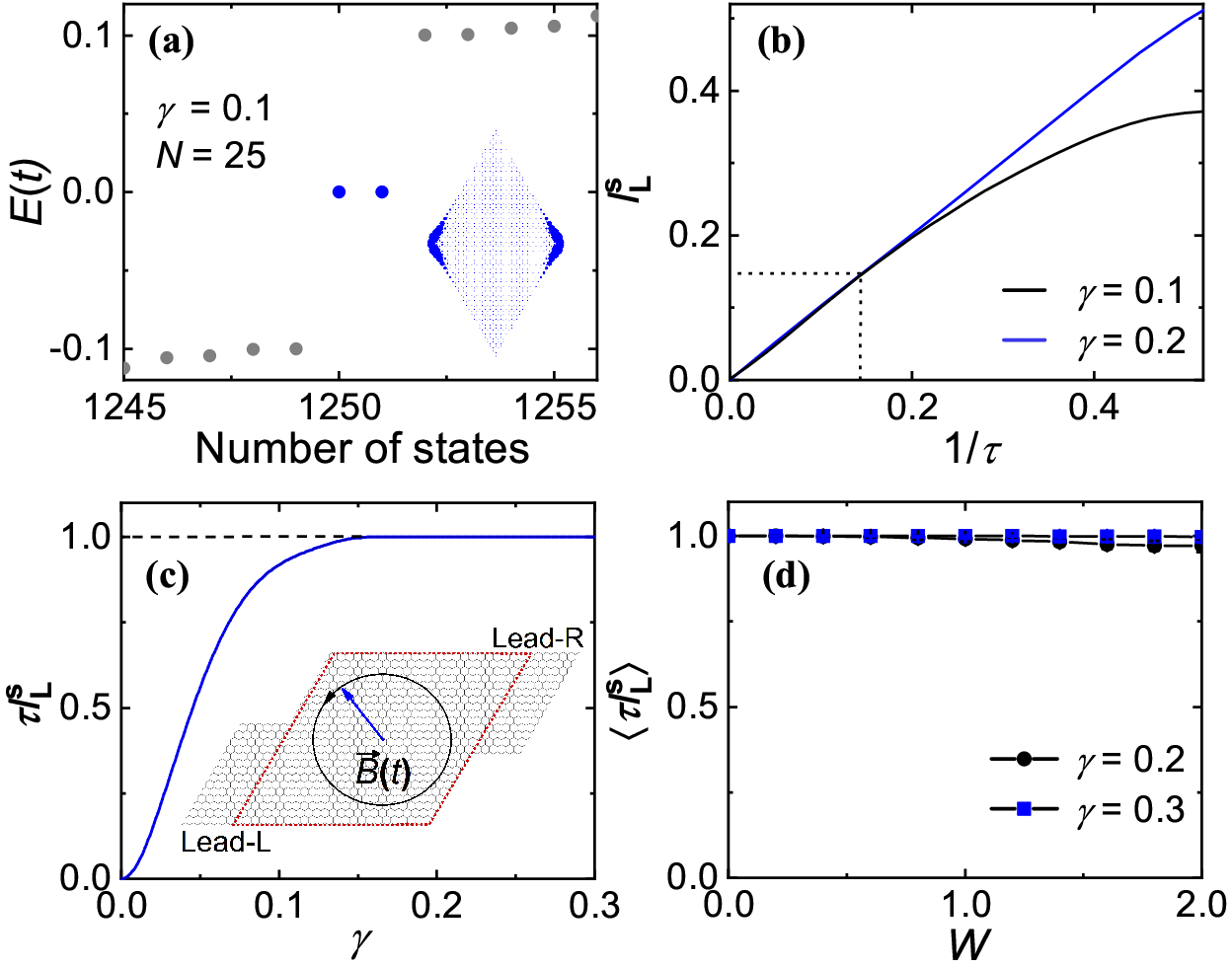}
\caption{(a) Zero energy levels and eigenfunction distribution of the Kane-Mele model of SOTI for a diamond-shaped honeycomb lattice. (b) Spin current as a function of the pumping frequency. (c) Total spin current per cycle versus the magnetic field strength. (d) Ensemble-averaged total spin current $\langle \tau I^s_L \rangle$ versus the disorder strength. $N=25$ in all panels. }\label{fig3}
\end{figure}

Disorder effect on the quantized spin current is also evaluated. Fig.~\ref{fig2}(c) shows that the ensemble-averaged total spin current $\langle \tau I^s_L \rangle$ is robust against Anderson-type disorder in a large range of disorder strength. As an important parameter, the Fermi energy is scanned. In the inset of Fig.~\ref{fig2}(c), $\tau I^s_L = 1$ is found in a large energy interval, which corresponds to the energy gap $E_g$ of a finite system with the same size. Meanwhile, this energy range for $\tau I^s_L = 1$ coincides with the linear regime for $I^s_L$ in Fig.~\ref{fig1}(d), which establishes the relation $E_g \approx \hbar \omega_c$ with $\omega_c$ the cutoff pumping frequency of the linear regime. This is another evidence that links quantized spin current in open systems with the topological property of finite systems. All these results demonstrate that, the quantized spin current $\tau I^s_L $ is universally independent of system parameters including the magnetic field strength, pumping frequency, system size, Fermi energy (within the topological energy gap), and robust against disorder, which represents a universal quantized transport signature of the BHZ model of SOTI.

\begin{table*}[tp]
    \centering
    \setlength\tabcolsep{1pt}
    \caption{Summary of the symmetry, topological, and spin pump features for different models. '$\surd$' ('$\times$') denotes the presence (absence) of the corresponding property.}
    \begin{tabular*}{\linewidth}{@{\extracolsep{\fill}} c|cccccc}
        \hline
        Model & BHZ & Kane-Mele & Pseudo spin & MTI & \makecell{Electric quadrupole\\insulator} & 3D model \\
        \hline
        TRS & $\surd$ & $\surd$ & $\surd$ & $\times$ & $\surd$ & $\surd$ \\
        First-order TI & Helical edge & Helical edge & Helical edge & Chiral edge & $\times$ & Helical surface \\
        Second-order TI & Corner & Corner & Corner & Corner & Corner & Hinge states \\
        Quantized spin pump & $n=1$ & $n=1$ & $n=2$ & $\times$ & $\times$ & $n=2$ \\
        \hline
    \end{tabular*} \label{table1}
\end{table*}

The quantized spin current $\tau I^s_L$ has clear physical meaning. Since $I^s_L$ is pure spin current and $\tau$ is the pumping period, $\tau I^s_L = n$ corresponds to integer spin quanta pumped per cycle,
\begin{equation}
\tau I^s_L = \frac{2 \pi}{\omega} I^s_L = \frac{2 \pi}{\omega}[ I_{L, \uparrow \uparrow} - I_{L, \downarrow \downarrow} ] = n. \nonumber
\end{equation}
We have shown that $\tau I^s_L$ is universally quantized to $n=1$ for the BHZ model. Different from quantized charge conductance of Chern insulators and quantized spin-degenerate charge conductance of Z$_2$-invariant TIs, quantized spin pump characterizes the universal transport signature of SOTI. It is reasonable since charge conductance is driven by electric field, while rotating magnetic field drives spin current. The validation of this quantized spin pump is further verified in other SOTI models, as shown below.

\section{The Kane-Mele model of SOTI}
The Kane-Mele model describes quantum spin Hall effect in 2D honeycomb lattice, such as graphene\cite{Kane} and silicene\cite{Yao11}. Recently, Jacutingaite family materials (Pt$_2$HgSe$_3$ and Pd$_2$HgSe$_3$) have been proposed as promising Kane-Mele quantum spin Hall materials with a predicted topological gap up to $0.5 ~eV$\cite{Marzari18,Tamai20,Nemes20,Wang22}. When introducing an in-plane magnetic field in the Kane-Mele model, SOTI emerges in 2D honeycomb lattice\cite{Ren}. The Kane-Mele Hamiltonian is expressed as
\begin{equation}
H_0=-t\sum_{\langle i j\rangle} c_i^{\dagger} \sigma_0 c_j+i t_{\text{SO}} \sum_{\langle\langle i j\rangle\rangle}\nu_{ij}c_i^{\dagger}\sigma_z c_j.
\end{equation}
Here $t$ is the nearest-neighbor hopping energy. $t_{\text{SO}}$ is the strength of intrinsic spin-orbit coupling between next-nearest-neighbor sites. $\nu_{ij}= +1 (-1)$ when an electron jumps from site $j$ to site $i$ by taking a clockwise (anticlockwise) turn. $H_{rot}$ due to the rotating magnetic field is the same as Eq.~(\ref{Hrot}).

The Kane-Mele model of SOTI is represented by a pair of topological corner states residing in a diamond-shape confined system, and the corresponding zero energy levels are inside a large gap, as shown in Fig.~\ref{fig3}(a). Pumped spin current $I^s_L$ of a two-lead system increases linearly with the pumping frequency in the region enclosed by dash lines in Fig.~\ref{fig3}(b). We observe the relation $E_g \approx \hbar \omega_c$ again, where $E_g$ is the band gap of the finite system and $\omega_c$ is the cutoff frequency of the linear pumping region. The total spin current $\tau I^s_L$ per cycle increases with the magnetic field strength in Fig.~\ref{fig3}(c), and quantizes to $n=1$ when $\gamma > 0.15$. Fig.~\ref{fig3}(d) demonstrates the robustness of quantized spin current against disorder. Other system parameters including system size and Fermi energy are also checked, which all support that spin current $\tau I^s_L$ is universally quantized for the Kane-Mele model of SOTI.

\section{Discussion and summary}
The spin pump properties via rotating magnetic field for other SOTI models are also investigated, including the pseudo spin model\cite{Song}, electric quadrupole insulator\cite{Benalcazar}, magnetic topological insulator (MTI)\cite{Ezawa18}, and 3D model with topological hinge states\cite{Schindler}. Although possessing different symmetries and first-order TI precursors, they all belong to the SOTI family. The main features are summarized in Table.~\ref{table1}, and detailed numerical results are presented in the {\color{blue}Supporting Information}.

Observations from Table.~\ref{table1} are in order:

(1) SOTI is induced by the in-plane magnetic field in three types of systems, which are categorized into: (a) Z$_2$-invariant first-order TI which has TRS; (b) electric quadrupole insulator (2D SSH chain) which maintains TRS; (c) magnetic topological insulator (MTI) that breaks TRS;

(2) Rotating the in-plane magnetic field pumps quantized spin current or integer spin quanta in SOTIs whose precursor is Z$_2$-invariant first-order TI; for the pseudo spin model and the 3D model, the quantization number is $n=2$ due to the degeneracy in the Hamiltonian; for electric quadrupole insulator and MTI, only spin-polarized charge currents can be generated;

(3) For the pseudo spin model and the 3D model of SOTI, the rotating magnetic field acts on pseudo spin, i.e., orbital degree of freedom, which shows that quantized pseudo spin pump is also accessible; in these two models, real spin $\sigma_z$ is not conserved but pseudo spin $\tau_z$ is conserved in $H_0$ part of the system Hamiltonian; the rotating magnetic field is coupled to pseudo spin $\tau_{x/y}$ and manipulates the strength and phase of nearest-neighbor hoppings; details are shown in the {\color{blue}Supporting Information};

(4) The 3D model of SOTI has conducting 1D hinge states as the second-order topological phase; quantized spin pump can be realized by connecting 2D leads to the 3D central region, where the way of connecting leads to the central region can be arbitrary; details are shown in the {\color{blue}Supporting Information};

(5) The quantized spin pump is independent of system size, Fermi energy, magnetic field strength, pumping frequency, spatial dimensionality, and robust against disorder, showing a universal feature.

In conclusion, for SOTIs evolved from Z$_2$-invariant TIs by applying an in-plane magnetic field to break time-reversal symmetry, quantized spin pump is achieved via rotating this magnetic field. The quantization of pumped spin current corresponds to the formation of topological corner states representing SOTI. Quantized spin pump is verified in various 2D and 3D SOTI models whose precursor is Z$_2$-invariant TI, and it is robust against disorder and independent of system parameters. As the universal quantized transport signature, quantized spin pump characterizes the bulk-boundary-transport correspondence of SOTIs.

The existence of higher-order bulk-boundary correspondence and 1D hinge modes has been experimentally confirmed in bismuth\cite{Neupert18,Madhavan21}. It was shown both theoretically and experimentally that spin pumping mechanism via rotating magnetic field enhances the Gilbert damping in ferromagnetic film/normal metal bilayers\cite{SpinPumpExpt,BauerPRL02,BauerPRB02}. Quantized spin pump can be realized by combining the SOTI states with the rotating magnetic field mechanism. We notice that higher-order topological phases are more easily fabricated in photonic crystals\cite{Dong19,Chen19,MLi20,photonicK} and acoustic materials\cite{Zhang19,Jia21,Qiu22,Liu23}. Specifically, it was found that acoustic higher-order topology can be generated from first-order with Zeeman-like fields\cite{Chen2022}. Since the quantized spin pump is also achievable via manipulating the pseudo spin such as the orbital degree of freedom, we expect that this quantization phenomenon can be observed in the second-order photonic and acoustic systems.

In the application aspect, such a quantized spin pump based on SOTI precisely pumps integer spin quanta. An optimal quantum parametric pump is noiseless\cite{Avron1,WangBG4} and was predicted to pump quantized charge per pumping period\cite{Levinson,WangBG5}. Based on 1D TI, Fu and Kane proposed a $Z_2$ adiabatic spin pump preserving TRS\cite{Kane06PRB}, which could pump nonquantized spin per cycle. Before the emergence of SOTI, X.-L. Qi et.al suggested that, when applying an in-plane magnetic field on one particular edge of the QSH state, a fractional charge $e/2$ is confined on the magnetic domain wall and rotating this magnetic field would generate a quantized dc electric current\cite{fraccharge}. The universally quantized spin pump proposed in this work can serve as an accurate, stable, and robust single-spin source, which is on demand in quantum computing and quantum information.

This work is supported by the National Natural Science Foundation of China (Grants No.~12034014, No.~12174262, and No.~12147164).

\end{document}